\newif\ifsingle
\newif\ifFullversion
\Fullversiontrue % comment out for short version

\ifsingle
\documentclass[11pt,draftclsnofoot, onecolumn]{IEEEtran}		
\else		
\documentclass[9pt,final, twocolumn]{IEEEtran}
\fi

\usepackage{times}
\usepackage{amsmath,dsfont}
\usepackage{amssymb,amsthm}
\usepackage{epsfig,verbatim}
\usepackage{subcaption}
\usepackage{setspace}
\usepackage{color}
\usepackage{cite}
\usepackage{epstopdf}
\usepackage{graphics}
\usepackage{accents}
\usepackage{acronym}
\usepackage[bookmarks,colorlinks]{hyperref}
\usepackage{booktabs}
\usepackage{mathtools}
\usepackage{enumitem}
\usepackage{multirow}
\usepackage{booktabs}
\usepackage{hyperref}
\usepackage{epstopdf} %converting to PDF

\usepackage[ruled,linesnumbered]{algorithm2e}  %,algochapter
\SetKwInput{KwData}{\textbf{Init}} 
\usepackage{algpseudocode} 
\newcommand{\myVec}[1]{{\boldsymbol{#1}}}

\newcommand{\mySet}[1]{\mathcal{#1}}
\definecolor{NewColor}{rgb}{0,0,0} %{0.2,0,0.5}

\ifsingle

\else

\fi % ------------------------------------------

\acrodef{adc}[ADC]{analog-to-digital convertor}
\acrodef{cs}[CS]{compressed sensing}
\acrodef{dtft}[DTFT]{discrete-time Fourier transform}
\acrodef{dnn}[NN]{neural network} 
\acrodef{csi}[CSI]{channel state information}
\acrodef{map}[MAP]{maximum a-posteriori probability}
\acrodef{snr}[SNR]{signal-to-noise ratio}
\acrodef{bs}[BS]{base station} 
\acrodef{iot}[IOT]{Interent of Things}
\acrodef{mimo}[MIMO]{multiple-input multiple-output}
\acrodef{mse}[MSE]{mean-squared error}
\acrodef{pdf}[PDF]{probability density function}
\acrodef{rv}[RV]{random variable}
\acrodef{ml}[ML]{machine learning}
\acrodef{mf}[MF]{matched filter}
\acrodef{fec}[FEC]{forward error correction}
\acrodef{rs}[RS]{Reed-Solomon}
\acrodef{lti}[LTI]{linear time-invariant}
\acrodef{wss}[WSS]{wide-sense stationary}
\acrodef{psd}[PSD]{power spectral density}
\acrodef{ser}[SER]{symbol error rate} 
\acrodef{ber}[BER]{bit error rate} 
\acrodef{sgd}[SGD]{stochastic gradient descent} 
\acrodef{isi}[ISI]{intersymbol interference}  
\acrodef{awgn}[AWGN]{additive white Gaussian noise} 
\acrodef{ut}[UT]{user terminal} 
\acrodef{mmw}[mmWave]{millimeter wave}
\acrodef{noma}[NOMA]{non-orthognal multiple access}
\acrodef{mac}[MAC]{mulitple access channel}
\acrodef{fl}[FL]{federated learning}
\acrodef{ct}[CT]{continuous-time}

\IEEEoverridecommandlockouts

\title{Federated mmWave Beam Selection Utilizing LIDAR Data}

	\author{ Mahdi Boloursaz Mashhadi$^*$, Mikolaj Jankowski$^*$, Tze-Yang Tung, Szymon Kobus, Deniz G\"{u}nd\"{u}z\\Dept. of Electrical and Electronic Eng., Imperial College London, UK
	\thanks{$^*$ Indicates equal contribution.\\This work was supported by the European Research Council (ERC) through project BEACON (grant no 677854).}}

\vspace{-0.75cm}

\begin{document}
	
	\maketitle
	\pagestyle{empty}
	\thispagestyle{empty}
	
%%%%%%%%%%%%%%%%%%%%%%%%%%%%%%%%%%%%%%%%%%%%%%%%%%
\begin{abstract}
	%%%%%%%%%%%%%%%%%%%%%%%%%%%%%%%%%%%%%%%%%%%%%%%%%%
	Efficient link configuration in millimeter wave (mmWave) communication systems is a crucial yet challenging task due to the overhead imposed by beam selection. For vehicle-to-infrastructure (V2I) networks, side information from LIDAR sensors mounted on the vehicles has been leveraged to reduce the beam search overhead. In this letter, we propose a federated LIDAR aided beam selection method for V2I mmWave communication systems. In the proposed scheme, connected vehicles collaborate to train a shared neural network (NN) on their locally available LIDAR data during normal operation of the system. We also propose a reduced-complexity convolutional NN (CNN) classifier architecture and LIDAR preprocessing, which significantly outperforms previous works in terms of both the performance and the complexity.
	
\end{abstract}
%%%%%%%%%%%%%%%%%%%%%%%%%%%%%%%%%%%%%%%%%%%%%%%%%%

%%%%%%%%%%%%%%%%%%%%%%%%%%%%%%%%%%%%%%%%%%%%%%%%%%%%%%
\vspace{-0.2cm}
\section{Introduction}\label{sec:intro}
\vspace{-0.1cm}
%%%%%%%%%%%%%%%%%%%%%%%%%%%%%%%%%%%%%%%%%%%%%%%%%%%%%%

Millimeter wave (mmWave) is a promising technology for high data rate vehicle-to-infrastructure (V2I) communications. However, efficient beam selection in mmWave communications is challenging due to the overhead imposed by the beam search process. Recently it was shown that side information from sensors mounted on vehicles can be exploited to reduce the beam-selection overhead for mmWave links. \color{black}For instance, position information can be used to query the most prominent mmWave beams \cite{1}. Inertial sensors placed on vehicle's antenna arrays enable efficient antenna element configuration by tracking the orientation of the vehicle \cite{brambilla2019inertial}. Furthermore, position and motion information can be jointly processed to further reduce the alignment overhead \cite{8277251}. From the infrastructure side, a radar located at the base station (BS) can help estimate the direction of arrival and aid the beam search \cite{7}. Spatial information obtained from out of band measurements is exploited in \cite{5, 6, Sub6G1, Sub6G2} where \cite{Sub6G1, Sub6G2} used sub-6GHz channel measurements to train neural network (NN)s for mmWave beamforming. Vision-aided approaches are considered in \cite{vision, alrabeiah2020millimeter, Alkhateeb}. BSs equipped with cameras are proposed to employ computer vision and deep learning techniques to predict mmWave blockage and beam strength in  \cite{alrabeiah2020millimeter}. The authors in \cite{Alkhateeb} build a panoramic point cloud from images taken within the cellular coverage area. This point cloud is then input to a neural network (NN) to predict the optimal beams.\color{black}

%The position information from vehicles is used in \cite{1, 2, 3, 4}, while out-of-band measurements are used in \cite{5, 6} for efficient mmWave beam selection. Information from a radar located in the infrastructure is shown to be beneficial for mmWave link establishment in \cite{7}. 

%LIDAR uses a laser to scan the environment and generates a three-dimensional (3D) image with pixels indicating relative positions from the sensor \cite{10}. Data from the LIDAR sensors mounted on vehicles can be exploited for improved beam-selection in mmWave V2I communications. On the other hand, the lack of analytical models that can relate LIDAR outputs to mmWave channels motivates employing a neural network (NN)-based approach to this problem. 

The use of light detection and ranging (LIDAR) technology is considered in \cite{8, 9}, where a NN architecture is trained over simultaneous LIDAR and ray-tracing channel datasets to identify $K$ beam directions that include the beam pair with the best channel condition between the vehicle and the BS with the highest probability. The approach in \cite{8, 9} is distributed, in the sense that, each vehicle uses the trained NN on the measurements from its own LIDAR sensor to infer its top-$K$ beam directions. It is shown in \cite{9} that such a distributed approach outperforms centralized beam selection, where a NN at the BS infers the best beams for all the vehicles in its coverage area either by combining LIDAR data from all the vehicles or using a single LIDAR sensor mounted at the BS. Although the NN performs beam selection inference in a distributed fashion in \cite{8, 9}, it is trained offline on LIDAR and channel measurements from all the vehicles gathered in a centralized dataset. However, in practice, gathering a large centralized dataset of individual LIDAR measurements from vehicles in each individual cell is challenging as it requires communicating a large amount of LIDAR point cloud data over the uplink channel. %Note also that a separate NN needs to be trained for the coverage area of each BS as the trained NN is site-specific, and will not perform well even for the same site if significant changes occur in the scattering environment. Therefore, continuous recollection of up-to-date LIDAR data and retraining or fine-tuning the NN weights is necessary during normal operation of the system. This means that a centralized approach imposes a continuous overhead on the system for transmitting up-to-date LIDAR measurements to the BS.      

%a LIDAR sensor at t It can be shown that performance of a centralized beam selection scheme carried out at the base station (BS) significantly degrades as the number of unconnected vehicles increases and hence, a distributed beam selection scheme at the vehicles is generally more favorable. The authors in \cite{9} showed that even in the all connected scenario, distributed beam selection outperforms its centralized counterpart specially in the non-line-of-sight (NLOS) cases. 

%\begin{figure}
%    \centering
%    \includegraphics[scale=.35]{images/FEDBeam1.jpg}
%    \caption{The proposed federated LIDAR-based beam selection scheme.}
%    \label{fig:scheme}
%\vspace{-0.1cm}
%\end{figure}

\color{black}This paper builds upon the unpublished work of the authors that recently won the ``AI/ML in 5G" competition ranking second in ``ML for mmWave beam selection" challenge organized by the International Telecommunications Union (ITU) \cite{challenge, ranking}.\color{black} In this work, we propose fully a distributed LIDAR-aided beam selection method for V2I mmWave communication systems, in which both the inference and training of the NN are performed in a distributed fashion at the vehicles in the coverage area of the BS. We propose a three-phase procedure, which enables the vehicles to periodically collect up-to-date data and train or fine-tune the NN in a federated manner during normal operation of the system. After the training phase, each vehicle leverages the trained NN and its locally available LIDAR data to infer a subset of beams that are most likely to contain the best transmitter/receiver beam pair. We also propose a reduced-complexity convolutional NN (CNN) architecture along with LIDAR preprocessing, which significantly outperforms previous works. The proposed architecture achieves a top-$10$ classification accuracy of $91.17\%$ on the benchmark Raymobtime dataset \cite{raymobtime}, which is a significant improvement over the previous works in \cite{8, 9}, while reducing the number of floating point operations (FLOPs) and parameter complexity of the NN by factors of 100 and 55, respectively. For further reproduction of the reported results, our codes are available at: \url{https://github.com/galidor/ITU_Beam_Selection_TF} %Federated training helps avoid the large communication overhead that would be imposed by the transmission of LIDAR measurements from the connected vehicles to the BS to gather a centralized dataset for offline training. 
%The reduction in the number of trainable NN parameters facilitates efficient federated training of the proposed architecture during normal operation of the system with reduced communication overhead.  

%The rest of the paper is organized as follows: Section~\ref{sec:Model} presents the system model. Section~\ref{sec:Approach} presents our proposed federated LIDAR-aided beam selection scheme. Simulation results are presented in Section~\ref{sec:results}. Finally, Section~\ref{sec:Conclusions} concludes the paper. For further reproduction of the reported results, our codes are available at: \url{https://github.com/galidor/ITU_Beam_Selection_TF}

%%%%%%%%%%%%%%%%%%%%%%%%%%%%%%%%%%%%%%%%%%%%%%%%%%%%%%
\vspace{-0.2cm}
\section{System Model}\label{sec:Model}
\vspace{-0.1cm}
%%%%%%%%%%%%%%%%%%%%%%%%%%%%%%%%%%%%%%%%%%%%%%%%%%%%%%
We consider a downlink orthogonal frequency division multiplexing (OFDM) mmWave system, where a BS located on the street curb serves connected vehicles in its coverage area over $N_c$ subcarriers. The BS and the vehicles are equipped with $N_t$ and $N_r$ antennas, respectively. Denote by $\mathbf{H}_n$ the downlink channel matrix from the BS to a vehicle over the $n$'th subcarrier. We assume that both the BS and the vehicle have antenna arrays with only one radio frequency (RF) chain and apply analog beamforming. We assume fixede beam codebooks $\mathcal{C}_t = \{\mathbf{f}_i\}_{i=1}^{C_t}$ and
$\mathcal{C}_r = \{\mathbf{w}_j\}_{j=1}^{C_r}$ at the transmitter and receiver sides, respectively. 

Utilizing a pair $(i, j) \in \mathcal{C}_t \times \mathcal{C}_r$ of precoder and combiner vectors, the resulting channel gain at subcarrier $n$ is $\mathbf{w}_j^H \mathbf{H}_n\mathbf{f}_i$, where $(\cdot)^H$ denotes the conjugate transpose. \color{black}For fair comparison with previous works \cite{8, 9}, we also assume a noise-free mmWave setting.\color{black}   For the $(i,j)$ pair, the  sum  power  gain  over all subcarriers is given by $y_{ij} =\sum_{n=1}^{N_c} |\mathbf{w}_j^H \mathbf{H}_n\mathbf{f}_i|^2$.
%\begin{align}
%    y_{ij} =\sum_{n=1}^{N_c} |\mathbf{w}_j^H \mathbf{H}_n\mathbf{f}_i|^2.
%\end{align}
Hence, the optimum beam label is $b^*=(i^*,j^*)=\underset{(i,j)}{\operatorname{argmax}}\  y_{ij}$. Without any side information, the transmitter and receiver would search through all $C_tC_r$ beam pairs to identify $b^*$. Our goal is to infer a small subset of $K$ beam pairs $\mathcal{S} = \{(i_k, j_k)\}_{k=1}^K \subset \mathcal{C}_t \times \mathcal{C}_r$ using the available position and LIDAR data, such that $b^* \in \mathcal{S}$. This results in a reduction of $\frac{K}{C_t \times C_r}$ in the search space for beam selection. In the next section, we propose a novel NN architecture as well as a federated training approach for top-$K$ beam classification from simultaneous position and LIDAR data.

\begin{figure}
    \centering
    \includegraphics[width=\linewidth]{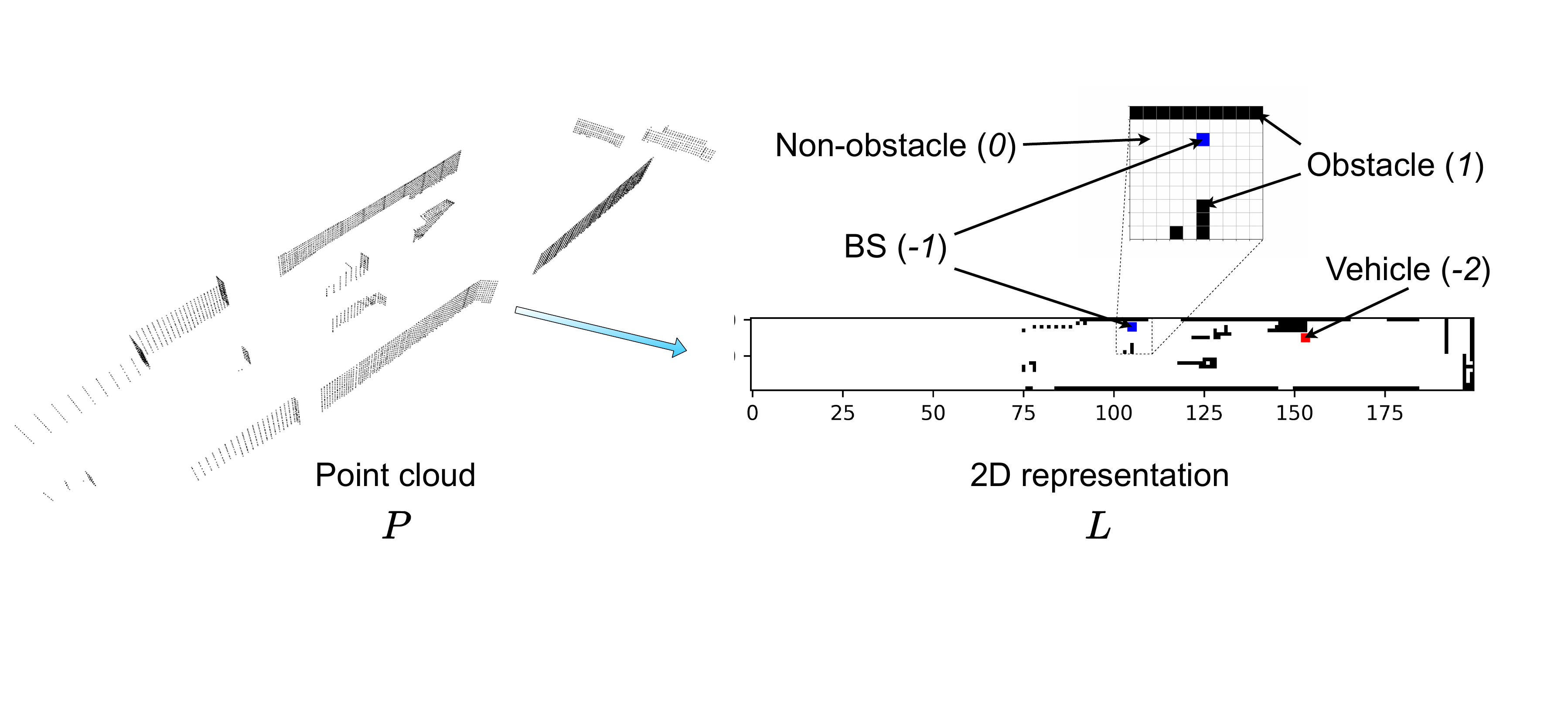}
    \vspace{-1.2cm}
    \caption{\color{black}Preprocessing of the LIDAR point cloud.}
    \label{fig:pcl22d}

\end{figure}

\begin{figure}
    \centering
    \includegraphics[width=\linewidth]{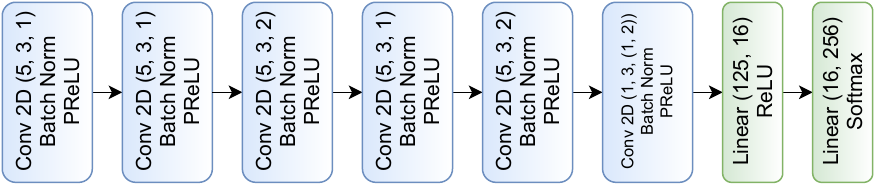}
    \caption{The proposed model architecture.}
    \label{fig:2dmodel}
%\vspace{-0.1cm}
\end{figure}

%%%%%%%%%%%%%%%%%%%%%%%%%%%%%%%%%%%%%%%%%%%%%%%%%%%%%%
\vspace{-0.2cm}
\section{Federated Beam Selection Utilizing LIDAR Data}\label{sec:Approach}
%\vspace{-0.1cm}
%%%%%%%%%%%%%%%%%%%%%%%%%%%%%%%%%%%%%%%%%%%%%%%%%%%%%%
We propose a novel data-driven beam selection scheme, where connected vehicles in the coverage area of a BS collaborate to train a shared NN for top-$K$ beam classification using their position and LIDAR data in a distributed manner. Collaborative training is orchestrated by the BS, and takes place during normal operation of the network.% as depicted in Fig. \ref{fig:scheme}. 

\vspace{-0.2cm}
\subsection{Three-Phase Network Operation}
\label{subsec:FE}
\vspace{-0.1cm}

Our proposed solution consists of three network operation phases: (i) data collection phase, (ii) federated training phase, and (iii) distributed inference phase. 

During phase (i), a subset of connected vehicles in the coverage area of the BS, denoted by $\mathcal{V}=\{v\}_{v=1}^{V}$, each acquires a local dataset $\mathcal{D}_v = \{(\mathcal{P}_v, \mathcal{B}_v)\}_{v \in \mathcal{V}}$, where $\mathcal{P}_v=\{P_i\}_{i=1}^{|\mathcal{D}_v|}$ contains instances of the point cloud $P_i$ recorded by the LIDAR sensor and $\mathcal{B}_v=\{b_i^*\}_{i=1}^{|\mathcal{D}_v|}$ contains the corresponding best beam pair labels $b_i^* \in \mathcal{C}_t \times \mathcal{C}_r$, i.e., index of the best beam pair. During this phase the vehicles employ any beam selection or tracking technique of their choice to identify the best beam pair. %Although brute-force beam search in this phase imposes an overhead on the network, it provides accurate beam labels required for training of the NN. 

%Define the $i$'th scene $(P_i, b_i^*)$ as the status of the environment captured by the LIDAR output $P_i$ and the corresponding beam label $b_i^*$, and denote the time period (in seconds) between consecutive scenes by $T_s$. Then, the total duration (in seconds) of phase (i) is $T_1=|\mathcal{D}_v| \times T_s$, and the total amount of data collected by all the vehicles for training of the NN is $|\mathcal{D}_U|=V \times |\mathcal{D}_v|= V \times \frac{T_1}{T_s}$. 

During phase (ii), vehicles with local datasets collaborate in a federated learning scheme to train or fine-tune a shared NN for top-$K$ beam classification. In particular, the vehicles employ federated averaging (FedAvg) \cite{mcmahan2016communication}, where a global model is sent to the vehicles by the BS at each round, and the vehicles perform mini-batch stochastic gradient descent (SGD) updates based on their local datasets. The local updates are aggregated by the BS, and used to update the global model for the next round. The duration of this phase is proportional to the  number of global aggregation rounds required to train the model, denoted by $N_a$. Note that, the vehicles train a single site-specific NN, which learns the statistical characteristics of the coverage area of the BS for efficient beam selection.

Finally, in phase (iii), any vehicle in the coverage area of the BS utilizes the most up-to-date trained/tuned NN on its local LIDAR data to infer $K$ beams and reduce the beam search overhead. Note that, in this phase, the BS can use a low frequency control channel to transmit the trained NN model to any new vehicle entering its coverage area. 

Each BS in a large network can orchestrate training of a site-specific NN for its own coverage area following the above three phases.  \color{black}Note that the above operation may either be initialized randomly, or from a NN pre-trained on a centralized dataset offline. As the above three phases take place periodically, the NN can automatically adapt to changes in the statistics of data over time through periodic fine-tuning of the NN with up-to-date data. Also note that the above phases take place during normal operation of the vehicular network and impose no interruption. The vehicles keep communicating with the BS as they would do without the above phases (utilizing any beam selection or tracking technique of their choice) but just record and collect the beam labels and the corresponding LIDAR inputs for training. Only some vehicles with sufficient dataset would transmit model updates when fine-tuning the NN is required. Once the NN is trained up to a desired accuracy, it is used for beam selection thereby significantly reducing the beam search space.\color{black}%The temporal efficiency of the proposed scheme is hence $\eta=\frac{T_3}{T_1+T_2+T_3}$. This three-phase process continues periodically to enable updating the NN parameters to adapt to the changes in the environment. We investigate the above points by simulations later in Section IV.

	\begin{algorithm} [t] 
		\caption{FedAvg for LIDAR-assisted beam selection}
		\label{alg:Algo1}
		\KwData{Initial parameters $\myVec{\theta}_v^{(0)} = \myVec{\theta}^{(0)}$, $\forall v \in \mathcal{V}$. }
		\For{each $m = 1,2,\ldots$  }{
			{Each vehicle performs a local epoch using mini-batch gradient decent iterations according to \eqref{eqn:SGD}} \;
			\If{$m$ is an integer multiple of $N_v$}
			{ Each vehicle $v$ sends $\myVec{g}_v^{(m)} = \myVec{\theta}_v^{(m)}-\myVec{\theta}_v^{(m-N_v)}$ to BS\;
			 BS computes $\myVec{\theta}^{(m)} =\myVec{\theta}^{(m-N_v)} +  \frac{\mu}{|\mathcal{V}|}\sum\myVec{g}_v^{(m)} $ \;
			 \label{stp:DownTrans} BS distributes $\myVec{\theta}^{(m)}$ such that $\myVec{\theta}_v^{(m)} = \myVec{\theta}^{(m)}$, $\forall v \in \mathcal{V}$\;
			}
		} 
		\KwOut{Trained $\myVec{\theta}^{(m)}$ shared among all vehicles.}
	\end{algorithm}

\begin{table*}
\centering
\caption{Comparison between the proposed NN architecture and the baseline in \cite{8, 9}, both trained in a centralized manner.}
\begin{tabular}{|c|c|c|c|c|} 
\hline
Model & Top-$10$ accuracy & Top-$10$ throughput ratio & FLOPs & \# of NN parameters, $|\myVec{\theta}|$ \\ 
\hline
\hline
Proposed centralized & $91.17 \pm 0.28 \%$ & $94.78 \pm 0.61\%$ & $1.72 \times 10^6$ & $7462$ \\
\hline
Baseline \cite{8, 9} & $83.92 \pm 0.93\%$ & $86.15 \pm 0.82\%$ & $179.01 \times 10^6$ & $403677$ \\
\hline
\end{tabular}
\label{tab:results}
\vspace{-0.2cm}
\end{table*}

%-----------------------------------
%	Model-Based Symbol Detection for Fading Channels
%----------------------------------- 
\vspace{-0.2cm}
\subsection{LIDAR and Location Preprocessing}
\label{subsec:FE}
\vspace{-0.1cm}

\color{black}For each scene, the LIDAR sensor mounted on each vehicle outputs a point cloud $P=\{(x_p, y_p, z_p)\}_{p=1}^{|P|}$, representing obstacles measured by the LIDAR sensor. Each vehicle $v$ also has its own location information $(x_v, y_v, z_v)$, and the BS location $(x_{BS}, y_{BS}, z_{BS})$, which is broadcast to all the vehicles. We preprocess this data to obtain a tensor of fixed size, which contains both the location and LIDAR data and is input to the NN for each scene. To reduce both the NN dimension and the computation load, we propose a two-dimensional (2D) representation of the LIDAR measurements, where we partition the coverage area of the BS into a grid of equal-size square cells from the top view (see Fig. 1). We define the corresponding 2D tensor $L$, where the cells containing the BS and the vehicle are set to -1 and -2, respectively, while each of the remaining cells is populated with a 1 if it accommodates at least one of the cloud points, and with a 0 otherwise. We remark that this 2D representation discards the height data along the z-axis, resulting in a significant reduction in the input size, and hence, the complexity of the NN, which in turn reduces the communication overhead for federated training.  \color{black} %Fig. \ref{fig:pcl22d} illustrates this preprocessing scheme. Moreover, we observed through experiments that the proposed 2D representation even improves the performance in comparison with a 3D representation.

%-----------------------------------
%	\FadeNet Architecture
%----------------------------------- 
\vspace{-0.2cm}
\subsection{NN Architecture}
\label{subsec:Architecture}
\vspace{-0.1cm}

Our NN architecture consists of 6 convolutional layers followed by 2 linear layers as depicted in Fig. \ref{fig:2dmodel}. In the convolutional layers, we vary the value of stride between 1 and 2, depending on whether we intend to downscale the intermediate features, or not. We apply batch normalization and parametric rectified linear unit (PReLU) activation after each convolutional layer. The first linear layer is followed by rectified linear unit (ReLU) activation, and softmax is used at the output to obtain the predictions.  \color{black}To achieve better generalization, convolutional layers downscale the features and ensure that only essential information is preserved. This helps avoid overfitting to the training data \cite{pooling_overfitting}. \color{black} Note that, to reduce the communication overhead for federated training, we have minimized the trainable model parameters utilizing a convolutional structure with limited kernel sizes. We denote the NN model function by $\pi(L;\myVec{\theta})$, which is a vector of length $C_t C_r$ at the softmax output. $L$ denotes the preprocessed LIDAR and location input, while $\myVec{\theta}$ denotes the trainable NN parameters. The best beam is predicted as $\hat{b^*}=\mathop{\arg \max}\limits_{b \in \mathcal{S}} y_b$, where the prediction set $\mathcal{S}$ is given by the top-$K$ softmax outputs.

%As we show later by simulations, our NN architecture significantly reduces both the flop counts and the number of trainable parameters while improving the performance compared to \cite{8, 9}.

%-----------------------------------
%	Distributed Training
%----------------------------------- 
\vspace{-0.2cm}
\subsection{Federated Training}
\label{subsec:Training}
\vspace{-0.1cm}
%Due to the individual characteristics of a specific vehicle (e.g., its trajectory, dimensions, speed, etc.), its local dataset may not capture all the subtleties of the coverage area. In such cases, the NN trained solely on a local dataset $\mathcal{D}_v$ is highly biased and may not operate reliably for other vehicles entering the coverage area of the BS. We exploit the fact that, while each vehicle may capture a limited amount of training data that is biased towards its own specific circumstances, the overall dataset captured by several vehicles (i.e., $\mathcal{D}_U=\{\mathcal{D}_v\}_{v=1}^{V}$) within the coverage area of the BS is more diverse to allow training a generalizable NN model for the coverage area in consideration. On the other hand, 

Gathering a large dataset of LIDAR measurements from various vehicles for centralized training at the BS imposes significant communication overhead, particularly due to the large size of LIDAR point cloud measurements. Instead, we propose a federated learning approach, where the vehicles collaborate to train a single NN architecture using their local datasets \cite{mcmahan2016communication}. To train our NN we use the empirical cross entropy loss, hence the local loss calculated at vehicle $v$ is given by 
%We use a training mechanism, which consists of $N_v$ local epochs using mini-batch SGD iterations carried out at each vehicle and $N_a$ aggregation rounds by the BS.  

\begin{equation}
    \psi_v(\myVec{\theta}, \mySet{D}_v) = -\frac{1}{|\mySet{D}_v|}\sum_{i=1}^{|\mathcal{D}_v|}\log [\pi\left(L_i; \myVec{\theta}\right)]_{b_i^*},
    \label{eqn:NetLoss}
\end{equation}
where $[\pi]_b$ denotes the $b$'th element of the model's softmax output. Each connected vehicle performs mini-batch SGD iterations to update its local vector of model parameters, denoted by $\myVec{\theta}_v$, via
\begin{equation}
    \label{eqn:SGD}
    \myVec{\theta}_v^{(l)} = \myVec{\theta}_v^{(l-1)} - \rho_l\nabla \psi_v(\myVec{\theta}_v^{(l-1)}, \{(b_{i_l},L_{i_l})\}_{i_l \in 1, \hdots, |\mySet{D}_v|}),
\end{equation}
where $l$ is the local iteration index, $\rho_l > 0$ is the local step-size, and the set $\{(b_{i_l},L_{i_l})\}_{i_l \in 1, \hdots, |\mySet{D}_v|}$ is a mini-batch of the local dataset with $i_l \in 1, \hdots, |\mySet{D}_v|$. The training consists of $N_v$ local epochs at each vehicle (i.e. $N_v$ cycles of training on the vehicle's local dataset) and $N_a$ aggregation rounds at the BS as summarized in Algorithm~\ref{alg:Algo1}.

Such distributed learning orchestrated by the BS during phase (ii) requires the vehicles to periodically exchange and synchronize their local model parameters $\myVec{\theta}_v$ through reliable low-rate communications with the BS. This imposes an overhead of communicating $O_{UL}=V\times N_a \times |\myVec{\theta}|$ float32 variables in the uplink and $O_{DL}=N_a \times |\myVec{\theta}|$ in the downlink channel. Minimizing the number of trainable parameters $|\myVec{\theta}|$ is hence critical to reduce the communication overhead during phase (ii) of the network operation.

%%%%%%%%%%%%%%%%%%%%%%%%%%%%%%%%%%%%%%%%%%%%%%%%%%%%%%
\vspace{-0.2cm}
\section{Numerical Evaluations}\label{sec:results}
\vspace{-0.1cm}
%%%%%%%%%%%%%%%%%%%%%%%%%%%%%%%%%%%%%%%%%%%%%%%%%%%%%%

We provide numerical evaluations on the benchmark Raymobtime datasets \cite{raymobtime}, where we train the models on samples from dataset s008 and test on those from s009. \color{black}Our training dataset includes 6482 line-of-sight (LOS) and 4712 non-line-of-sight (NLOS) samples, while our test dataset includes 1473 LOS and 8165 NLOS samples, respectively (refer to \cite{raymobtime, raymobtime1} for details on these datasets e.g. locations, frequencies, etc.).\color{black} For performance comparison, we use the top-$K$ classification accuracy defined as the probability of correctly identifying the optimal beam pair within the top-$K$ output of the network, and the top-$K$ throughput ratio, $R$, defined as $R\triangleq(\sum_{t=1}^{T} \log_2(1+y_{\Tilde{i}\Tilde{j}}))/(\sum_{t=1}^{T} \log_2(1+y_{i^*j^*}))$,
%\begin{equation}
%    \label{eqn:throughput}
%    R\triangleq\frac{\sum_{t=1}^{T} \log_2(1+y_{\Tilde{i}\Tilde{j}})}{\sum_{t=1}^{T} \log_2(1+y_{i^*j^*})},
%\end{equation}
where $T$ is the number of test samples, and $(i^*, j^*)$ and $(\Tilde{i}, \Tilde{i})$ denote the optimum beam pair index and the best beam pair within the top-$K$ prediction set $\mathcal{S}$, respectively. 

In Table \ref{tab:results}, we compare the performance of the NN architecture presented in Subsection \ref{subsec:Architecture} with the baseline architecture proposed in \cite{8, 9}, both trained in a centralized manner. In this experiment, we trained our model using the Adam optimizer \cite{kingma2014adam} with an initial learning rate of $10^{-3}$ and batch size of 16, and train the models for 20 epochs. The grid dimensions to generate features in Section III.B is set to $20\times200$. In Table \ref{tab:results}, we present 95\% confidence intervals for Top-$10$ accuracy and throughput ratio of the models calculated from 10 Monte Carlo simulations. %Besides the learning rate adjustment imposed by the Adam optimizer, we further reduce the learning rate by a factor of 10 after the 10th epoch. 

According to Table \ref{tab:results}, our proposed architecture not only outperforms those in \cite{8} and \cite{9} in terms of both the top-$10$ accuracy and the throughput ratio, but also significantly reduces the complexity of the model. Our architecture reduces the FLOPs and the number of trainable parameters roughly by factors of 100 and 55, respectively. Such a significant reduction in the number of trainable model parameters is specifically desirable in federated training as it leads to a significant reduction of the communication overhead.

\begin{figure}
    \centering
    \includegraphics[scale=.7]{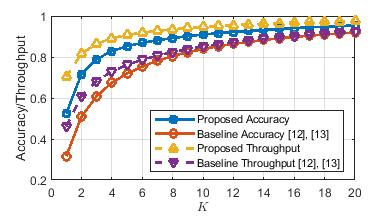}
    \caption{Top-$K$ accuracy and throughput ratio as a function of $K$.}
    \label{fig:Acc_Thr}
%\vspace{-0.1cm}
\end{figure}

Remember that the beam search complexity of these schemes depends on $K$, the size of the prediction set. Figure \ref{fig:Acc_Thr} plots the top-$K$ accuracy and throughput ratio for the proposed and baseline architectures as a function of $K$, when trained in a centralized fashion. It is observed that, our proposed model architecture significantly outperforms \cite{8, 9}, e.g., to achieve a throughput ratio $R \geq 90 \%$, our proposed model architecture requires $K \ge 3$ while the baseline needs $K \ge 16$. This is more than 5 times reduction in the required search space for beam selection. Also, the proposed architecture can achieve close to $80 \%$ of the optimal throughput with $K=1$; that is, with no beam search at all.

\color{black}Note that the results reported in Table I and Fig. 4 are the average values achieved on both LOS and NLOS samples. However, the NLOS case is more challenging due to blockages of the rays by other vehicles/objects. When measuring the Top-$10$ accuracy of our proposed approach separately on LOS and NLOS samples, we get $94.50 \%$ and $90.77 \%$, respectively, which shows that the proposed approach performs very well on NLOS samples as well and significantly outperforms previous works \cite{8, 9}.\color{black}

\color{black}We also compare the performance of the proposed approach with a NN-based approach that only assumes access to the location data. Our NN architecture for the location-only approach is a 4-layer fully connected NN with 16, 32 and 16 nodes in the hidden layers, each followed by ReLU activation and batch normalization. This architecture has been hand-crafted for the best performance, and takes as input the relative location of the vehicle with respect to the BS. It achieves top-$10$ accuracy and throughput ratio of $87.48 \%$ and $91.96 \%$, respectively. Hence, the performance is significantly improved when the LIDAR data is used, which is expected, as the LIDAR data gives a better understanding of the scene and includes information on objects or obstacles that cause blockage of the mmWave rays.\color{black}

We next evaluate the performance of our proposed federated beam selection scheme. To generate the local dataset at each connected vehicle $v$, we choose $|\mathcal{D}_v|=11000/V$ samples from the training set s008 uniformly at random, where $11000$ is the total number of samples in s008. We use mini-batch SGD with an initial learning rate of $0.2$ and exponential rate decay of $0.001$ with a batch size of 16 for local optimization at the vehicles. We set the learning rate $\mu=0.2$ for aggregation at the BS. We provide the performance of our proposed federated beam selection scheme in Table \ref{tab:Perf}. \color{black}Here, we start training from a randomly initialized global model. \color{black} The notation $(N_a)^{0.88}$ in this table represents the number of global aggregation rounds required for the training to achieve a top-$10$ accuracy larger than $88 \%$. This is an important measure as it determines the communication overhead required to train the model to the specified accuracy. Notations $(O_{DL})^{0.88}$ and $(O_{UL})^{0.88}$ used in this table represent this overhead in terms of the number of float32 variables needed to be communicated over the downlink and uplink channels, respectively. According to Table \ref{tab:Perf}, the number of aggregation rounds required to achieve top-$10$ accuracy larger than $88 \%$ increases when more vehicles take part in federated training. A larger $(N_a)^{0.88}$ increases the communication overhead. However, thanks to our simple NN architecture, which only has $|\myVec{\theta}|=7462$ trainable parameters, the maximum communication overhead required for federated training (i.e., $1620 \times 7462 \sim 1.2\times10^7$ float32 communications for $V=20$, $N_v=1$) is orders of magnitude smaller than the overhead that would be imposed by offloading the LIDAR point clouds to the BS for centralized training (i.e., $\sim 4\times10^9$ float32 communications for samples in s008). \color{black}Note that the communication overhead for federated training can further be significantly reduced utilizing gradient quantization \cite{GQ1, GQ2, GQ3}, gradient sparsification \cite{GS1, GS2}, and over-the-air aggregation \cite{AirComp1, AirComp2} techniques.\color{black} Also note that utilizing various neural architecture search (NAS) approaches \cite{NAS1, NAS2, NAS3}, we can design further simplified but well-performing NN architectures, thereby reducing the communication overhead for federated training.% and this is a direction for future research.\color{black}

\color{black}Note that the size of the local dataset at a vehicle, i.e. $|\mathcal{D}_v|$, depends on how long the vehicle stays in the coverage area of the BS and how frequently it can collect samples. $|\mathcal{D}_v|$ is a design parameter to be set for each specific site depending on dimensions of the coverage area and the traffic flow speed. For the Raymobtime s008/s009 datasets used in our simulations, the coverage area is a $337\times202 \mathrm{m}^2$ region covering the intersection of Kent and 19th street in Rosslyn, Virginia, the average vehicle speed is $8.2 \frac{\mathrm{m}}{\mathrm{s}}$ and the sampling period is $0.1 \mathrm{s}$ [19]. Hence, an average vehicle traveling along the Kent street can collect approximately $\frac{337}{8.2\times0.1}\approx411$ samples during its stay in the coverage area of the BS, which is approximately the local dataset size used in Table II for $V=20$. Note that vehicles may collect more samples if they are parked, move slower, return to the cell multiple times, or have faster sampling equipment.\color{black}%, hence, improving the performance.  

%and duration of the training phase. On the other hand, the duration of the data collection phase decreases with $V$. This is because we keep the total number of samples the same across all cases, and hence, increasing $V$ means that each vehicle needs to collect fewer samples. Here, we assume that there are always $V$ vehicles in the cell that can participate in the training process. This leads to a tradeoff between the duration of the data collection and training phases, and the communication overhead. For practical deployment, the number of vehicles participating in federated training should be decided according to the requirements of the system and the amount of communication overhead that can be afforded. 

\color{black}The last column in Table \ref{tab:Perf} reports the final top-10 accuracy achieved for each number of vehicles $V$ and local epochs $N_v$. This column shows a slight performance degradation when more vehicles take part in federated training. This is due to the limited number of training samples available to each vehicle. Although we distribute the samples among vehicles uniformly at random, the local vehicle datasets are still slightly skewed due to the limited number of training samples (e.g., 11K samples available in s008). This leads to the catastrophic forgetting phenomenon \cite{CF1, CF2}, where increasing $N_v$ tends to overfit to local datasets, which may not efficiently represent the true distribution across the cell. This can be mitigated utilizing ideas similar to \cite{CF1, CF2}, and is a direction for future research.\color{black}

\begin{table}
\centering
\caption{\color{black}Performance of federated beam selection when initialized from an untrained global model.}
\begin{tabular}{|c|c|c|c|c|c|}
\hline
                   $V$ & $N_v$ & $(N_a)^{0.88}$ & $(O_{DL})^{0.88}$ & $(O_{UL})^{0.88}$ & Top-$10$ Acc. \\ \hline\hline
\multirow{3}{*}{5}  & $1$ & $19$ & $19|\theta|$  & $95|\theta|$ & $90.12 \%$ \\ \cline{2-6} 
                    & $2$ & $13$ & $13|\theta|$ & $65|\theta|$ & $90.34 \%$ \\ \cline{2-6} 
                    & $5$ & $10$ & $10|\theta|$ & $50|\theta|$ & $89.92 \%$ \\ \hline\hline
\multirow{3}{*}{10}  & $1$ & $31$ & $31|\theta|$  & $310|\theta|$ & $89.77 \%$ \\ \cline{2-6} 
                    & $2$ & $22$ & $22|\theta|$ & $220|\theta|$ & $89.16 \%$ \\ \cline{2-6} 
                    & $5$ & $15$ & $15|\theta|$ & $150|\theta|$ & $88.64 \%$ \\ \hline\hline
\multirow{3}{*}{20}  & $1$ & $81$ & $81|\theta|$ & $1620|\theta|$ & $88.81 \%$ \\ \cline{2-6}
                    & $2$ & $48$ & $48|\theta|$ & $960|\theta|$ & $88.53 \%$ \\ \cline{2-6} 
                    & $5$ & NA & NA & NA & $87.33 \%$ \\ \hline
\end{tabular}
\label{tab:Perf}
\end{table}

%Finally, we investigate the effect of the number of local epochs $N_v$, on top-$10$ classification performance in Table \ref{tab:EpochNum}. We fix $N_v \times N_a=20$ to ensure roughly the same amount of overall computations for training. While a larger $N_v$ reduces the total number of aggregation rounds and the corresponding communication overhead, it causes some performance reduction. This is because, with more local epochs, the locally trained models tend to overfit to the local datasets, which in turn hampers the convergence to an efficient global model and degrades the performance.

\begin{table}[]
\centering
\caption{\color{black}Performance of federated beam selection when initialized from an offline-trained model.}
\begin{tabular}{|c|c|c|c|c|c|}
\hline
{\color {black} }                     & {\color {black} }                     & \multicolumn{2}{c|}{{\color {black} 2K from s008 (Rosslyn)}}       & \multicolumn{2}{c|}{{\color {black} 2K from s007 (Beijing)}}       \\ \cline{3-6} 
\multirow{-2}{*}{{\color {black} $V$}}  & \multirow{-2}{*}{{\color {black} $N_v$}} & {\color {black} $(N_a)^{0.88}$} & {\color {black} Top-$10$ Acc.} & {\color {black} $(N_a)^{0.88}$} & {\color {black} Top-$10$ Acc.} \\ \hline\hline
{\color {black} }                     & {\color {black} $1$}                    & {\color {black} $17$}  & {\color {black} $90.29 \%$}   & {\color {black} $23$}  & {\color {black} $89.35 \%$}   \\ \cline{2-6} 
{\color {black} }                     & {\color {black} $2$}                    & {\color {black} $10$}  & {\color {black} $89.86 \%$}   & {\color {black} $14$}  & {\color {black} $88.73 \%$}   \\ \cline{2-6} 
\multirow{-3}{*}{{\color {black} $10$}} & {\color {black} $5$}                    & {\color {black} $6$}  & {\color {black} $89.31 \%$}  & {\color {black} $9$} & {\color {black} $88.19 \%$}  \\ \hline
\end{tabular}
\label{tab:Init}
\end{table}

% \begin{table}[]
% \centering
% \caption{\color{black}Perfromance comparison between the proposed model, when fed with features from \cite{8, 9}, and model from \cite{8, 9}, when fed with our 2D features.}
% \begin{tabular}{|c|c|c|c|c|}
% \hline
%  & \specialcell{Proposed \\ centralized} & \specialcell{\cite{8, 9} with \\ 2D features} & \specialcell{Proposed with \\ features from \cite{8, 9}} & Baseline \cite{8, 9} \\
% \hline
% Top-10 acc. & 91.17\% & 85.76\% & 84.86\% & 83.92\% \\
% \hline
% \end{tabular}
% \label{tab:FeaturesComparison}
% \end{table}

\color{black}Finally, in Table \ref{tab:Init}, we present the performance of our proposed federated beam selection scheme when initialized from an offline-trained model. We have considered two scenarios as below:

-The model is trained offline for 20 epochs on 2K samples taken randomly from Raymobtime s008 achieving an initial Top-$10$ accuracy of $81 \%$ on the test dataset s009. %This offline-trained model is then used as the initial global model for federated training. For federated training, the remaining $11-2=$9K samples of s008 are randomly distributed among the $10$ users participating in federated training. %The training parameters, i.e. optimizer, learning rate, batchsize, etc. remain the same as before.   

-The model is trained off-line for 20 epochs on 2K samples taken randomly from Raymobtime s007 achieving an initial Top-$10$ accuracy of $56 \%$ on the test dataset s009. 

In both cases, the offline-trained model is then used as the initial global model for federated training. For federated training, 9K samples from s008 are distributed randomly among the $10$ vehicles participating in federated training. %The training parameters, i.e. optimizer, learning rate, batchsize, etc. remain the same as before.

Referring to \cite{raymobtime, raymobtime1}, s007, s008 and s009 are Raymobtime datasets from different locations (i.e., s008 and s009 are from Rosslyn, while s007 is from Beijing) at the same 60GHz carrier frequency. The purpose of the above two scenarios is not only to study the effects of using an initial model trained offline, but also to investigate how well the model can adapt if significant changes happen in the coverage area of the BS. It can be concluded from the results presented in Table \ref{tab:Init} that using an offline trained NN generally reduces the required number of aggregation rounds (hence, reducing the communication overhead) as well as the number of samples required to be collected by the vehicles (here from 11K to 9K). If the dataset used for offline training is a good representative of the scattering environment, then a slight improvement in the accuracy is observed as well. But even if it is not, the federated training scheme can adapt the model to the new environment and still achieve a very good accuracy.\color{black}
%%%%%%%%%%%%%%%%%%%%%%%%%%%%%%%%%%%%%%%%%%%%%%%%%%%%%%
\vspace{-0.05cm}
\section{Conclusions}\label{sec:Conclusions}
%\vspace{-0.1cm}
%%%%%%%%%%%%%%%%%%%%%%%%%%%%%%%%%%%%%%%%%%%%%%%%%%%%%%
We have studied efficient link configuration in mmWave V2I communication networks, and considered exploiting side information in the form of LIDAR and position data in a supervised learning scheme to reduce the beam search overhead. In this letter, we first proposed LIDAR preprocessing and a convolutional NN architecture that improves the state-of-the-art classification accuracy with a significantly reduced model complexity. We have then proposed a federated training scheme that enables connected vehicles to collaboratively train a shared NN on their locally available LIDAR data. Once the NN is collaboratively trained, any vehicle entering the coverage area of the BS can employ it to reduce the beam search overhead.%Thanks to the simple NN architecture, this federated solution significantly reduces the communication overhead compared to the centralized approach, which requires communicating massive LIDAR measurements over the wireless link. 

%----------------------------------------------------------------------------------------
%	BIBLIOGRAPHY

\bibliographystyle{IEEEtran}
\bibliography{IEEEabrv,refs}
%----------------------------------------------------------------------------------------	

\end{document}